\documentclass[aps, prl, reprint, showpacs, twocolumn, 10pt, groupedaddress]{revtex4-1}

\usepackage{graphicx, amsmath, dsfont, dcolumn, bm, times, color, braket, amssymb, multirow}

\begin{document}

\bibliographystyle{apsrev4-2}

\title{Emergence of a condensate with finite-energy Cooper pairing in hybrid exciton/superconductor systems}

\author{Viktoriia Kornich}
\affiliation{Institute for Theoretical Physics and Astrophysics, University of W\"urzburg, 97074 W\"urzburg, Germany }

\date{\today}

\begin{abstract}
We study theoretically a setup consisting of excitons formed in two valleys, with proximity-induced Cooper pairing, different in conduction and valence bands. Due to the combination of Coulomb interaction with superconducting proximity effects, Cooper pairing between electrons from conduction and valence bands from different valleys is formed. This finite-energy intervalley Cooper pairing has both even- and odd-frequency contributions. We show that there is a phase transition into formation of a robust macroscopic condensate of such Cooper pairs and present characteristics of the corresponding Higgs modes.  
\end{abstract}

\maketitle

\let\oldvec\vec
\renewcommand{\vec}[1]{\ensuremath{\boldsymbol{#1}}}
{\it Introduction.--} Exciton is an electron-hole pair bound via Coulomb interaction. It is often formed in semiconductors, when an electron is excited from the valence band into the conduction band via light \cite{koch:natmater06}. Due to its structure, exciton is a charge-neutral particle and a composite boson. Both of these properties allow for interesting phenomena and applications \cite{mueller:2dma18}, e.g., Bose-Einstein and other types of condensation \cite{butov:jpcm04, wang:nature19}. Excitons can also effectively couple to light, producing exciton-polaritons that also exhibit condensation and superfluidity \cite{utsunomiya:natphys08, byrnes:prl10, lerario:natphys17}. 

Another well-known example of a composite boson is a Cooper pair. The conventional type of a Cooper pair is two electrons with opposite spins and momenta. However, there are cases, when electrons pair with non-zero net momentum, e.g., FFLO state \cite{fulde:pr64, larkin:zetp64}. The spins of electrons in a Cooper pair can form triplet in, e.g., p-wave superconductor \cite{oldeolthof:prl21}. If electrons have also non-zero orbital momenta, Cooper pairing can be classified by their total orbital momenta, giving, e.g., quintet or septet pairing \cite{brydon:prl16, bahari:prr22}. Cooper pair can be formed at finite energy, i.e. constituent electrons belong to different energy states, e.g., in an Ising superconductor  \cite{tang:prl21, hoerhold:2dmater23} or a conventional one \cite{chakraborty:prb22}. Our main interest is to find a system with robust and macroscopic condensate of finite-energy Cooper pairs, possibly of unconventional type of pairing in frequency domain, because odd-frequency superconductors exhibit various exotic properties, e.g., paramagnetic Meissner effect \cite{abrahams:prb95, lee:prb17, parhizgar:prb21}.

\begin{figure}[tb]
	\begin{center}
		\includegraphics[width=\linewidth]{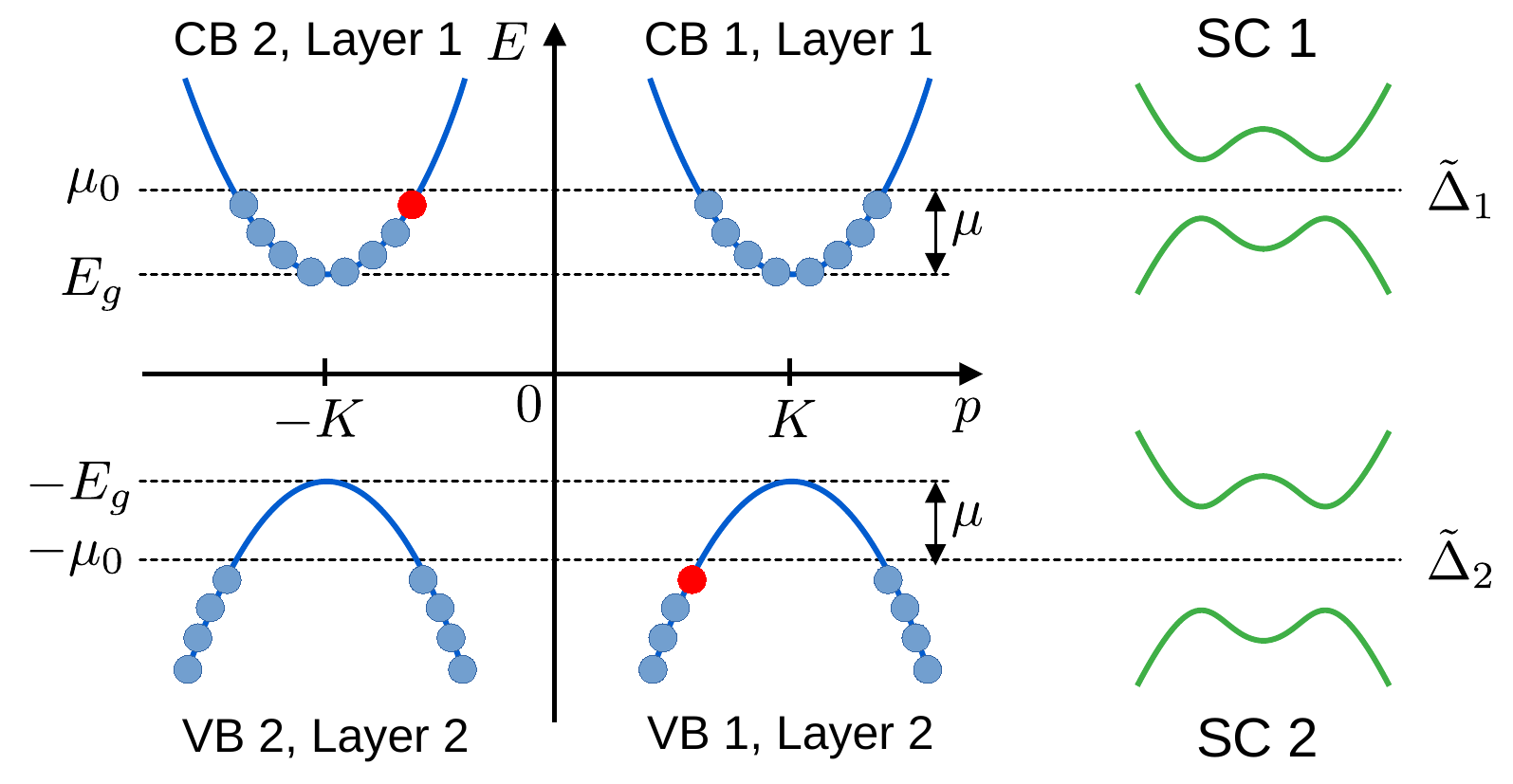}
		\caption{Schematic picture of the studied system: a semiconductor has two valleys at $\pm K$, with the gap $2E_g$. The chemical potential $\mu=\mu_0-E_g$ shows how the conduction band is filled by the pumped electrons, consequently, the valence band is emptied by the same amount (the occupied states are depicted with the blue circles). Electrons in the conduction band experience proximity effect from superconductor 1 (with the gap $2\tilde{\Delta}_1$) and electrons in the valence band from superconductor 2 (with the gap $2\tilde{\Delta}_2$). This can happen, if conduction and valence bands are also separated in real space, e.g. are situated in different crystal layers. One finite-energy intervalley Cooper pair is pictorially shown with red circles.}
		\label{fig:spectrum}
	\end{center}
\end{figure}

In this work, we study emergence of a condensate of Cooper pairs formed by electrons from conductance and valence bands from different valleys, i.e. finite-energy intervalley Cooper pairing. We consider two valleys with excitons, see Fig.~\ref{fig:spectrum}. The superconducting proximity effect can be induced in both conductance and valence bands: they both are not fully filled. In principle, the proximity effects induced in the conductance and valence bands can be different due to the initial coupling between superconductor and electrons from the conduction band and consequent coupling via spin-orbit interaction to electrons in the valence band \cite{moghaddam:prb14}. Even if the proximity effect in the valence band is induced by direct coupling between superconductor and the valence band electrons, it will be different from the one with the conduction band electrons due to the difference in the wavefunctions of electrons in conductance and valence bands. Nonetheless, we would like to assume that there are two superconductors that produce proximity effect in our system: superconductor 1 for conduction band and superconductor 2 for the valence band, see Fig.~\ref{fig:spectrum}.

We show that we obtain finite-energy intervalley Cooper pairing with even- and odd-frequency contributions from the combination of Coulomb interaction between electrons and holes, which form excitons, and different superconducting proximity effects in conduction and valence bands. We then study, whether the induced Cooper pairing leads to the actual phase transition with the macroscopic condensate. For that, we employ time-dependent Ginzburg-Landau theory and derive the spectrum of the corresponding Higgs modes. We present the phase diagram in the space of the proximity-induced superconducting mean fields. Importantly, the emergence of this finite-energy Cooper pair condensate is very robust at the phases of superconductors $\phi_1=\pi$ and $\phi_2=0$. Therefore we expect that this setup can be actually used for further theoretical and experimental investigation of finite-energy Cooper pair condensates and odd-frequency condensates. We discuss possible experimental realization of it in heterostructures with layers of transition metal dichalcogenides and in semiconducting quantum wells.

{\it Model.--} We consider a semiconductor with two valleys, symmetric with respect to the center of the Brillouin zone, with their extrema being at $\pm K$, see Fig. \ref{fig:spectrum}. Electrons are pumped into the conduction band from the valence band, and we assume that their lifetime is enough long to consider the effects we are interested in within the static formalism \cite{kamide:prl10, lee:prl18}. The pumping of electrons gives the effective chemical potential $\mu=\mu_0-E_g$, where $\mu_0$ is the chemical potential for conduction band electrons with respect to $E=0$, and $E_g$ is half of the semiconducting gap. The Hamiltonian for the valleys without intervalley interactions is

\begin{eqnarray}
\label{eq:H0}
\nonumber H_0=\sum_{j=1,2;k}c_{j,k}^\dagger \left(\frac{k^2}{2m}-\mu\right)c_{j,k}+v^\dagger_{j,k}\left(-\frac{k^2}{2m}+\mu\right)v_{j,k}\\ \nonumber +\sum_{k,k',q}\left[ \frac{V_q}{2}(c_{j,k+q}^\dagger c_{j,k'-q}^\dagger c_{j,k'} c_{j,k}+v^\dagger_{j,k+q}v^\dagger_{j,k'-q}v_{j,k'}v_{j,k})\right.\\ \left.+W_q c^\dagger_{j,k+q}v_{j,k'-q}^\dagger v_{j,k'}c_{j,k}\right].\ \ \ 
\end{eqnarray}
Here, operators $c_{j,k}$ and $v_{j,k}$ are annihilation operators of electrons in the conduction and valence bands of the valley $j$, respectively. The interaction potentials $V_q$ and $W_q$ are due to Coulomb interaction and are different, if conduction and valence bands are spatially separated, see e.g., Ref. \cite{danovich:prb18}. The momenta $k$ are counted from $\pm K$ for the first and the second valleys, respectively.

The terms with $V_q$ can be approximated within Hartree-Fock approximation and give a shift in energy, which we absorb into $E_g$. The term with $W_q$, though, should be considered in more detail, because Coulomb electron-hole interaction allows for formation of excitons, that can form excitonic condensate \cite{kogar:science17}. This condensate should be taken into account at least in the mean field approximation. For that, we perform Hubbard-Stratonovich transformation with the mean fields $\Delta_{j,k}$ and $\bar{\Delta}_{j,k}$ of the action associated with $H_0$ in Matsubara representation, and obtain:
\begin{eqnarray}
\label{eq:S0}
\nonumber \mathcal{S}_0=\sum_{j=1,2}\sum_{k,n}\left[ \bar{c}_{j,n,k}\left(-i\omega_n+\frac{k^2}{2m}-\mu\right)c_{j,n,k}\right.\\ \left. \nonumber+\bar{v}_{j,n,k}\left(-i\omega_n-\frac{k^2}{2m}+\mu\right)v_{j,n,k}-\bar{\Delta}_{j,k} \bar{v}_{j,n,k} c_{j,n,k}\right. \\ \left.-\Delta_{j,k} \bar{c}_{j,n,k} v_{j,n,k}+\beta L^2\bar{\Delta}_{j,k} W_{k-k'}^{-1}\Delta_{j,k'}\right].\ \
\end{eqnarray}
Here, $\omega_n$ is a Matsubara frequency, $\beta=1/T$, and $c$, $\bar{c}$, $v$, and $\bar{v}$ are Grassmann fields corresponding to operators $c$ , $c^\dagger$, $v$, and $v^\dagger$. We assume that the main contribution to excitonic condensate comes from electron-hole pairs with opposite momenta \cite{lozovik:prb02, byrnes:prl10, hanai:prb18}.

The intervalley Coulomb coupling is described as follows:
\begin{eqnarray}
\label{eq:IntervalleyInt}
\nonumber \sum_{k,k',q}V_{q}[c_{1,k+q}^\dagger c_{2,k'-q}^\dagger c_{2,k'}c_{1,k}+v_{1,k+q}^\dagger v_{2,k'-q}^\dagger v_{2,k'}v_{1,k}]\\ +W_q[c_{1,k+q}^\dagger v_{2,k'-q}^\dagger v_{2,k'}c_{1,k}+c_{2,k+q}^\dagger v_{1,k'-q}^\dagger v_{1,k'}c_{2,k}].\ \ \ \
\end{eqnarray}
The first two terms can be again approximated within Hartree-Fock approximation and give the shift of $E_g$. The Fock term should be smaller than in case of intravalley interaction because of the large momentum mismatch between the valleys, $2K$. 

The last two terms of Eq.~(\ref{eq:IntervalleyInt}) can potentially be represented in direct and exchange channels for Hubbard-Stratonovich transformation, i.e. with the mean fields related to $\langle c_{1,k+q}^\dagger c_{1,k}\rangle$, $\langle v_{2,k'-q}^\dagger v_{2,k'}\rangle$ and $\langle c_{1,k+q}^\dagger v_{2,k'}\rangle$, $\langle v_{2,k'-q}^\dagger c_{1,k}\rangle$, respectively. The mean fields from the direct channel can be absorbed into $E_g$, while the exchange channel could give non-zero mean field in case of intervalley excitons occurring due to Coulomb interaction. However, $W_{2K+k}\ll W_{k}$ and we consider valleys with the same dispersion, and therefore we neglect its contribution here. Thus, these terms do not give additional mean fields.

We place conventional s-wave superconductor layers in proximity with the layers with excitons: superconductor with the order parameter $\tilde{\Delta}_1e^{i\phi_1}$ near the layer with electrons (layer 1) and superconductor with the order parameter $\tilde{\Delta}_2e^{i\phi_2}$ near the layer with holes (layer 2), see Fig.~\ref{fig:spectrum}. Assuming that there is a sufficient insulating barrier between the layers 1 and 2, we obtain proximity-induced superconducting spectral gaps for electrons in the layer 1 and layer 2. The chemical potential of superconductors must be tuned with $\sim \pm\mu_0$ or at least $\sim \pm E_g$, so that superconducting pairing is induced in a band with electrons, not in the gap $2E_g$, where there are no charge carriers. In order to suppress possible supercurrent flowing through this structure, we consider $\phi_2-\phi_1=\{0,\pi \}$. The proximity effect should couple electrons with opposite momenta, i.e. electrons from valley 1 and valley 2, giving:
\begin{eqnarray}
\nonumber H_{\rm SC}=\sum_k(\Delta_1e^{i\phi_1}c_{1,k}^\dagger c_{2,-k}^\dagger+\Delta_1e^{-i\phi_1}c_{2,-k}c_{1,k}+\\ +\Delta_2 e^{i\phi_2}v_{1,k}^\dagger v_{2,-k}^\dagger+\Delta_2 e^{-i\phi_2}v_{2,-k}v_{1,k}),\ \ \ 
\end{eqnarray}
where $\Delta_1$ and $\Delta_2$ are different from $\tilde{\Delta}_1$ and $\tilde{\Delta}_2$, because they are induced due to the proximity effect, and thus depend on the wavefunction overlap between electrons in the excitonic layers and in the superconductors, and other parameters of the structure.

If we calculate the Green's function of this setup as $G_0=(i\omega_n-H_0^\Delta-H_{\rm SC})^{-1}$, where $H_0^\Delta$ contains mean fields $\Delta$ as in Eq.~(\ref{eq:S0}) instead of full Coulomb interaction from Eq.~(\ref{eq:H0}), we will obtain the term that couple $c_{1,k}^\dagger$ with $v_{2,-k}^\dagger$:
\begin{eqnarray}
\nonumber[\Delta(e^{i\phi_1}\Delta_1-e^{i\phi_2}\Delta_2)(E_k+i\omega_n)](E_k^4+\Delta^4+2\Delta^2\omega_n^2\\ \nonumber+(\Delta_1^2+\omega_n^2)(\Delta_2^2+\omega_n^2)+E_k^2(2\Delta^2+\Delta_1^2+\Delta_2^2+2\omega_n^2)\\ +2\Delta^2\Delta_1\Delta_2\cos{(\phi_1-\phi_2)})^{-1}.\ \ \ 
\end{eqnarray}
The term coupling $v_{1,k}^\dagger$ and $c_{2,-k}^\dagger$ has $-i$ in front of $\omega_n$ in the numerator. The corresponding terms between annihilation operators are Hermitian conjugate to these ones except for the term $i\omega_n$ in the numerator.
All these terms contain both even and odd in $\omega_n$ contributions. Thus, there is odd-frequency Cooper pairing at finite energy, i.e. between electrons in the conduction and valence bands, as a result of Coulomb interaction between electrons and holes, $\Delta$, and proximity-induced $\Delta_1 e^{i\phi_1}$ and $\Delta_2 e^{i\phi_2}$. It is important that $\Delta_1 e^{i\phi_1}$ and $\Delta_2 e^{i\phi_2}$ are different, otherwise there is no pairing. This can be also understood from the point of view that an ordered state is usually a consequence of symmetry breaking.

{\it Time-dependent Ginzburg-Landau theory.--} However, it is still unclear whether this pairing leads to a corresponding condensate. In order to clarify this question, we place mean fields $A$ denoting the finite-energy Cooper pairing in the Hamiltonian, and then study via time-dependent Ginzburg-Landau theory, if $A$ can become an order parameter, i.e. lead to the macroscopic condensate that is described by $A$. For simplicity of the calculation we assume that $W_{k}\approx W={\rm const}$, and then $\Delta\approx{\rm const}$. This is motivated by the fact that we consider small $k$ only, $k\ll K$, and $W_k$ is a smooth function of $k$ resulting from Coulomb potential in 2D and electron and hole wavefunctions of the Gaussian shape \cite{leon:prb01, kyriienko:prb12}.  In order to suppress possible Josephson current, we assume $\phi_1=\pi$ and $\phi_2=0$ or $\pi$, that can be regulated, e.g., by SQUID-type of device. Thus, the action is
\begin{eqnarray}
\mathcal{S}&=&\nonumber2\beta L^2 W^{-1}\Delta^2-\sum_{n,k}\bar{\Psi}_{n,k}G^{-1}\Psi_{n,k},\\
G^{-1}&=&\begin{pmatrix}i\omega_n-E_k & \Delta & A_{n,k} & \Delta_1 \\ \Delta & i\omega_n+E_k & -\Delta_2 e^{i\phi_2} & A_{-n,k}\\ A_{n,k} & -\Delta_2 e^{-i\phi_2} & i\omega_n-E_k & -\Delta\\ \Delta_1 & A_{-n,k} & -\Delta & i\omega_n+E_k\end{pmatrix},\ \ \ \ \ 
\end{eqnarray}  
where the inverse Green's function $G^{-1}$ is in the basis $\Psi_{n,k}=\{c_{1,n,k}, v_{1,n,k}, \bar{v}_{2,-n,-k}, \bar{c}_{2,-n,-k}\}$. We can integrate out these fermionic degrees of freedom in the partition function, obtaining
\begin{eqnarray}
\mathcal{Z}=\mathcal{Z}_0\int e^{-\mathcal{S}}\mathcal{D}[A, \Delta, \Psi, \bar{\Psi}]=\\ \nonumber=\mathcal{Z}_0\int \mathcal{D}[A, \Delta ]e^{-2\beta L^2W^{-1}\Delta^2+{\rm Tr}\ln{[G^{-1}]}}.
\end{eqnarray}
The effective action that we have obtained in the second line allows us to build Ginzburg-Landau theory \cite{altland:book10}.  We can expand 
\begin{eqnarray}
{\rm Tr}\ln[G^{-1}]={\rm Tr}\ln{G_0^{-1}}-\sum_{n=1}^\infty \frac{{\rm Tr}[(G_0H_A)^{2n}]}{2n},
\end{eqnarray}
where we have neglected all odd terms due to Gaussian integration in $\mathcal{Z}$, and $H_A$ is the $A$-dependent part of $G^{-1}$. The first term is an unperturbed one, but the sum gives the necessary expansion in Ginzburg-Landau functional.

\begin{figure}[tb]
	\begin{center}
		\includegraphics[width=0.9\linewidth]{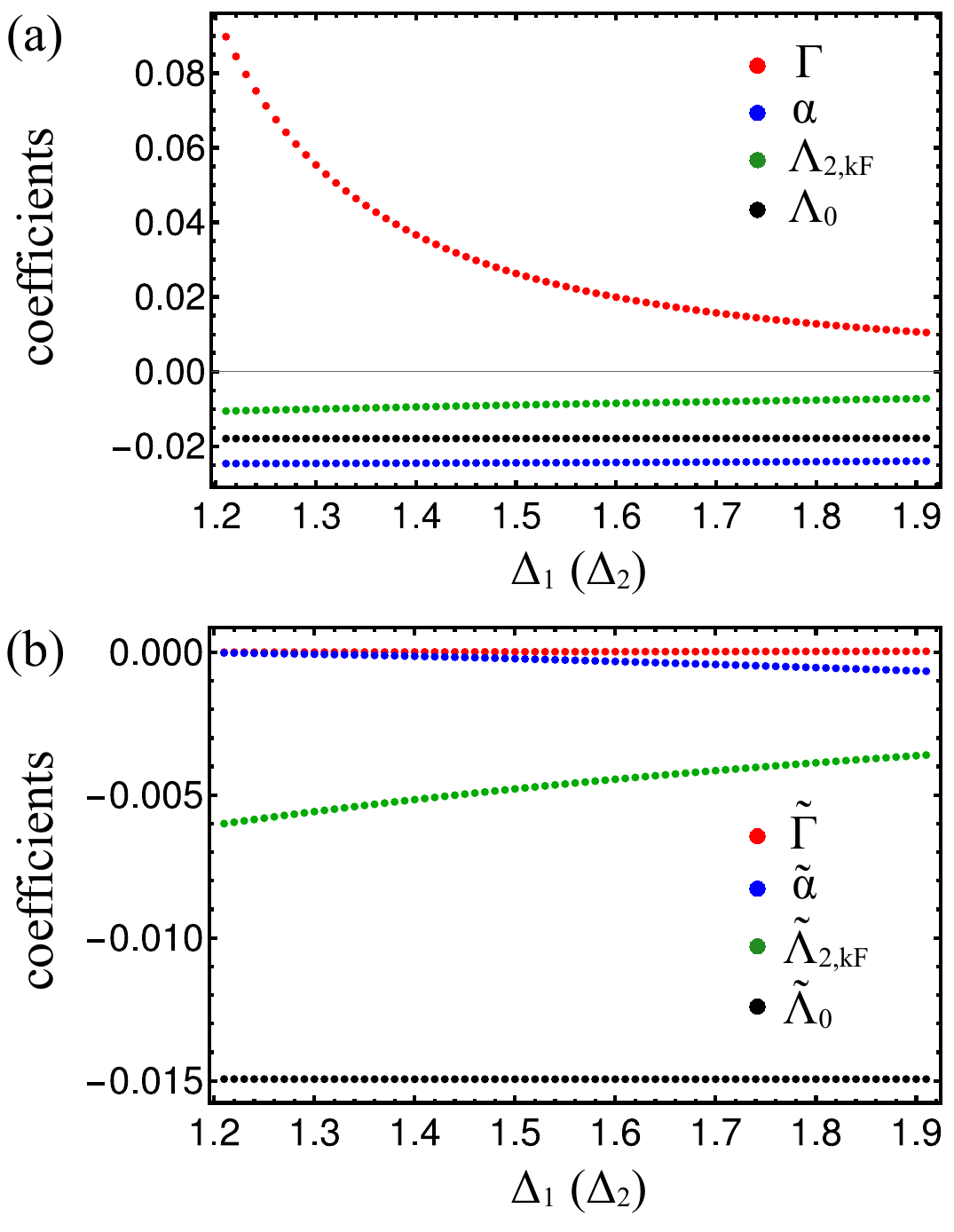}
		\caption{The dependence of $\Gamma$, $\alpha$, $\Lambda_{2,k_F}$, $\Lambda_0$, $\tilde{\Gamma}$, $\tilde{\alpha}$, $\tilde{\Lambda}_{2,k_F}$, and $\tilde{\Lambda}_0$ on $\Delta_1$ (or $\Delta_2$, because the dependences are the same). These coefficients constitute the dispersions of Higgs modes corresponding to the condensate of finite-energy Cooper pairs: (a) $\Re{[a_{m,q}]}$ and (b) $\Im{[a_{m,q}]}$. Here, we have taken $\Delta=1$, $m=1$, $\mu=2(|\Delta|+|\Delta_1|+|\Delta_2|)$, $\phi_2=0$, and $\Delta_2 ({\rm or} \ \ \Delta_1)=1.1$.}
		\label{fig:coefficients}
	\end{center}
\end{figure}

We assume for simplicity that frequency and momentum dependences of $A_{n,k}$ are weak, taking into account that $A_{n,k}$ has both even and odd in frequency parts, and there is no strongly momentum-dependent interaction that would produce mean field $A$. Thus, they can be taken into account just in the second order of expansion:
\begin{eqnarray}
\mathcal{S}^{(2)}=\frac{1}{2\beta L^2}\sum_{n,m;k,q}G_{0,n,k}H_{A,m,q}G_{0,n-m,k-q}H_{A,-m,-q}.\ \ \
\end{eqnarray}
As the system is uniform in 2D plane, we consider the time and space degrees of freedom independently, i.e., with $H_{A,m}$ and $H_{A,q}$. We sum over Matsubara frequencies $\omega_n$ and then expand in $\omega_m$ and $q-k$, respectively, up to the second order. We integrate over energy $E_k$ the terms of the series in $\omega_m$, and we also integrate over $k$ the term of the zero-approximation in $q-k$. We obtain
\begin{eqnarray}
\nonumber\mathcal{S}_{\rm eff}=\{\Lambda_0+\sum_{k,q}\frac{\Lambda_{2,k}}{L^2}(q-k)^2+\sum_m\Gamma\omega_m^2+\alpha\}(\Re [A_{m,q}])^2 
\\\nonumber+\{\tilde{\Lambda}_0+\sum_{k,q}\frac{\tilde{\Lambda}_{2,k}}{L^2}(q-k)^2+\sum_m\tilde{\Gamma}\omega_m^2+\tilde{\alpha}\}(\Im [A_{m,q}])^2\\ \label{eq:Seff}+bA_{0}^4+\gamma A_{0}^6.\ \ \ \ \ \ 
\end{eqnarray}
Thus, there are two non-degenerate Higgs modes: $\Re [A_{m,q}]$ and $\Im [A_{m,q}]$. We expand the action up to the sixth power in $H_A$, because we have found that $b<0$ in the large area of the phase space, therefore the sixth power in $H_A$ is necessary for stabilization of the action. Considering that frequency- and momentum-dependent part of $A$ is small, we define $A_{m,q}=A_0+a_{m,q}$, where $A_0$ is the equilibrium state of the mean field $A$, and $A_0^2=\frac{-b+\sqrt{b^2-3\alpha\gamma}}{3\gamma}$. The saddle point equations $\delta \mathcal{S}_{\rm eff}/\delta \Re[a_{l,p}]=\delta \mathcal{S}_{\rm eff}/\delta \Im[a_{l,p}]=0$ give the dispersions of the Higgs modes:
\begin{eqnarray}
\Gamma\omega^2=\sum_{k}\frac{\Lambda_{2,k}}{L^2}(p-k)^2+\Lambda_0+\alpha,\\
\tilde{\Gamma}\omega^2=\sum_{k}\frac{\tilde{\Lambda}_{2,k}}{L^2}(p-k)^2+\tilde{\Lambda}_0+\tilde{\alpha},
\end{eqnarray}
where we have transferred to the real frequency $\omega$ after variation of $\mathcal{S}_{\rm eff}$. The dependence of the coefficients $\Gamma$, $\alpha$, $\Lambda_0$, $\Lambda_{2,k_F}$, and $\tilde{\Gamma}$, $\tilde{\alpha}$, $\tilde{\Lambda}_0$, $\tilde{\Lambda}_{2,k_F}$ on $\Delta_1$ (or $\Delta_2$, as the dependences are the same) are shown in Fig.~\ref{fig:coefficients}. There, we have taken $\Delta=1$, $m=1$, and $\phi_2=0$. In order to be sure that the pumping of electrons is above possible interaction-induced gap, $\mu=2(|\Delta|+|\Delta_1|+|\Delta_2|)$. As $\Delta_1=\Delta_2=\Delta$ with $\phi_1=\pi$ and $\phi_2=0$ induces numerical instability at $E_k=0$, we take $\Delta_2=1.1$ ($\Delta_1=1.1$, if the dependence is on $\Delta_2$). In Fig.~\ref{fig:coefficients}~(a), we can see that $\Gamma$ has the most pronounced dependence on $\Delta_1$ and $\Delta_2$ compared to $\alpha$, $\Lambda_0$, and $\Lambda_{2,k_F}$. For $1.21<\Delta_1\lesssim1.4$, we have approximately $\Gamma\propto C_1+C_2\Delta_1^{-6}$, where $C_{1,2}$ are constants, for $1.4\lesssim\Delta_1<1.91$ the power-law is approximately $\Gamma\propto C_3+C_4\Delta_1^{-5}$ with $C_{3,4}$ being constants. The parameters of the Higgs mode $\Im{[A_{m,q}]}$ shown in Fig.~\ref{fig:coefficients}~(b) do not exhibit any strong dependences.

\begin{figure}[tb]
	\begin{center}
		\includegraphics[width=0.7\linewidth]{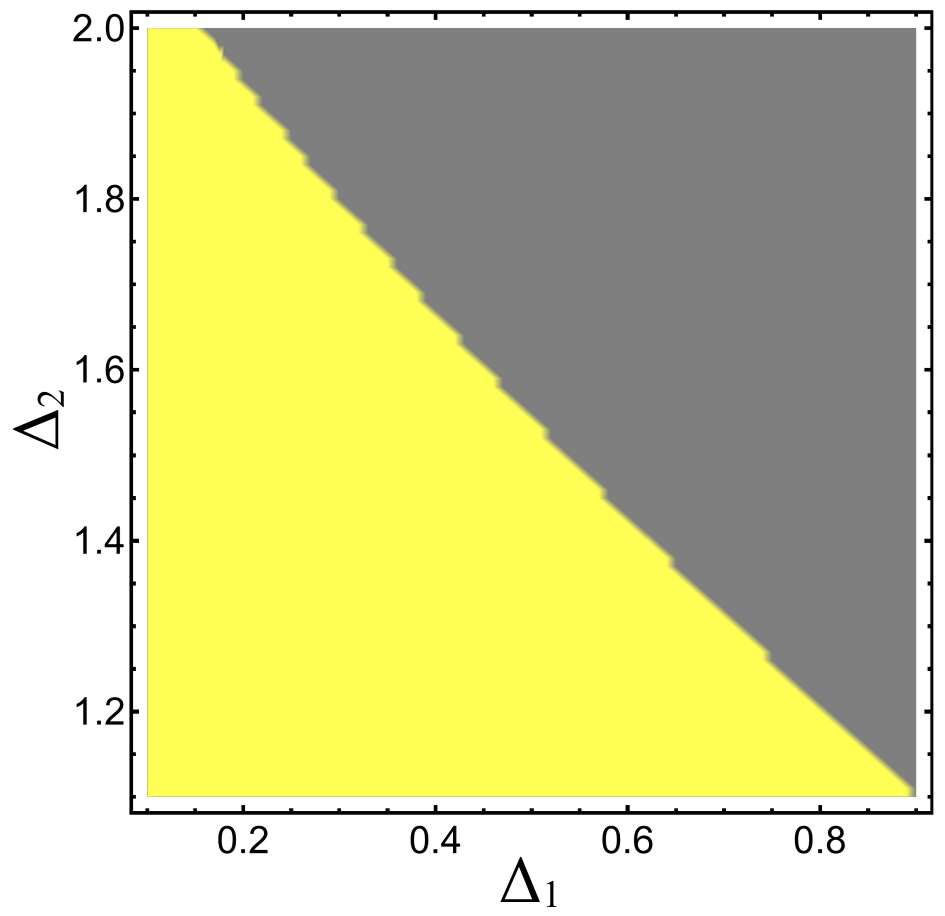}
		\caption{The phase diagram in space of proximity-induced $\Delta_1$ and $\Delta_2$. The yellow region denotes presence of the condensate of finite-energy intervalley Cooper pairs. The grey region denotes its absence in frames of our model. This is a density plot over discrete numerical data and we opted not to use interpolation methods. We can see that the phase region, where the condensate is present, is really wide. }
		\label{fig:phase_diagram}
	\end{center}
\end{figure}

We note that Higgs modes can be used in order to detect a condensate, e.g., via driving with the light, see Refs. \cite{schwarz:natcommun20, chu:natcommun20, matsunaga:prl13}, where the third harmonic generation can be a strong indicator of the presence of a Higgs mode of a certain order parameter.

{\it Phase transition.--} In order to study phase transition conditions, we neglect the dependence on frequency and momentum and keep only coefficient $\alpha$ for the second order of expansion in Eq.~(\ref{eq:Seff}) and correspondingly $A_0^2$ instead of $|A_{m,q}|^2$. We plot the phase diagrams in the axes $\Delta_1$ and $\Delta_2$, see Fig.~\ref{fig:phase_diagram}. We have used the following parameters: $\Delta=1$, $\mu=2(|\Delta|+|\Delta_1|+|\Delta_2|)$, $\phi_2=\pi$, and $\beta=100000$ to designate very low temperatures numerically. We have chosen $\phi_2=\pi$ here in order to show that the condensate can form also without phase difference between superconductors. In Fig.~\ref{fig:phase_diagram}, the yellow color denotes presence of finite-energy intervalley Cooper pair condensate. The grey color means that $\mathcal{S}_{\rm eff}(A_0)$ from Eq.~(\ref{eq:Seff}) does not give a stable condensate, but in principle we do not deny its presence completely, because maybe higher orders of $A_0$ could stabilize it. Our aim is to show that there is a large phase space in the coordinates $\{\Delta_1,\Delta_2\}$ that gives the condensate with Cooper pairing of electrons from different valleys and different energy bands. We would like to note that for the phases $\phi_1=\pi$ and $\phi_2=0$, the condensate occurs almost any enough large ($\gtrsim 0.1 \Delta$) $\Delta_1$, $\Delta_2$ that we have tried, meaning that this condensate is highly stable. The condensate is also present in certain ranges of $\Delta_1(\Delta_2)$ for $\Delta_2=0(\Delta_1=0)$, because one superconductor couples the valleys too.

\begin{figure}[tb]
	\begin{center}
		\includegraphics[width=0.7\linewidth]{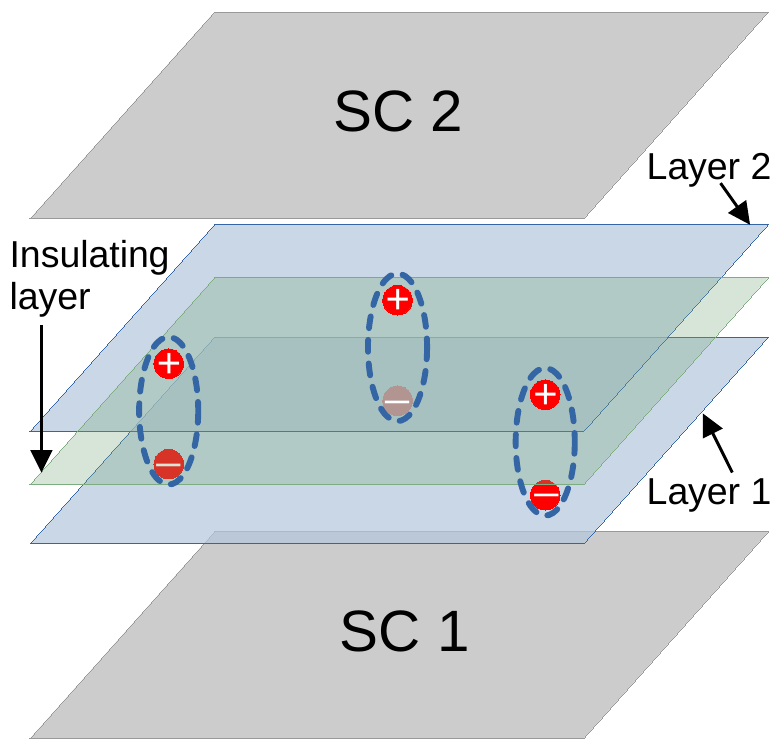}
		\caption{Possible setup. Two monolayers divided by an insulating layer. The monolayer 1 contains electrons and the monolayer 2 contains holes from externally-excited electron-hole pairs. Monolayers 1 and 2 experience superconducting proximity effects from the superconducting leads 1 and 2, respectively.}
		\label{fig:setup}
	\end{center}
\end{figure}

{\it Experimental realization.--} We think that the general scheme discussed above can be applied to bilayers of transition metal dichalcogenides (TMDs) or semiconducting quantum wells. The band valleys symmetric with respect to the center of the Brillouin zone is characteristic to TMD materials \cite{mak:natnanotechnol12}. In TMD heterostructures, electrons and holes can be localized in different layers \cite{fogler:natcommun14, shi:natnanotechnol22} and thus induce spatially indirect excitons, see Fig.~\ref{fig:setup}. The insulating layer in between TMD layers is usually made from hBN \cite{calman:natcommun18}. For the experimental setup closely related to Fig.~\ref{fig:spectrum} see, e.g., Ref. \cite{cai:nanolett24}, where the layers hosting excitons are WSe$_2$ and WS$_2$ with hBN barrier in between them.  The proximity-induced superconductivity has already been demonstrated experimentally on a MoS$_2$ monolayer \cite{trainer:acs20} and in a quantum spin Hall edge state of WTe$_2$ \cite{luepke:natphys20}. We note that indirect excitons can be induced in other heterostructures, e.g., GaAs quantum wells hosting electrons and holes with AlGaAs barrier \cite{hasling:apl15} or InAs quantum dot molecule with AlGaAs barrier in between the dots \cite{bopp:prb23}. Quantum well systems could be even better than TMD heterostructures due to their stability and more macroscopic size, but we need a semiconductor with two symmetric valleys, therefore GaAs quantum wells do not fit our purpose. Si/SiGe quantum wells have two lowest valleys, the difference to Fig.~\ref{fig:spectrum} is that the excitons will be indirect also in momentum space \cite{cheng:prb00}. However, we do not expect this to eliminate finite-energy Cooper pair condensate. Superconducting proximity effect can be obtained in quantum wells \cite{mayer:aem20}, and Si doped with Ga is known to be superconducting \cite{skrotzki:apl10}. In our opinion, both TMD heterostructures and semiconducting quantum wells are potentially suitable for experimental realization of the finite-energy Cooper pair condensate under study.

{\it Conclusions.}  We have shown theoretically, that the excitonic system with two valleys and proximity-induced superconducting pairing of electrons in a conduction and valence bands can have phase transition into the state with the condensate of Cooper pairs formed by electrons from different valleys and bands. This pairing happens due to Coulomb interaction between electrons and holes in excitons and proximity-induced electron pairing. The induced finite-energy Cooper pairing has both even- and odd-frequency components. We have derived the characteristics of the Higgs modes of this condensate using time-dependent Ginzburg-Landau theory, and proposed to use heterostructures of transition metal dichalcogenides or semiconducting quantum wells, as a possible platform for experimental realization of it. We believe that Cooper pair formation in complex band-valley structures can allow for methods of partial engineering of frequency dependence of unconventional Cooper pair condensates. 

\begin{acknowledgments}
	I acknowledge useful discussions with Sven H\"ofling, Sebastian Klembt, J\"urgen K\"onig, Atanu Patra, and Bj\"orn Trauzettel. This work was supported by the W{\"u}rzburg-Dresden Cluster of Excellence on Complexity and Topology in Quantum Matter (EXC2147, project-id 390858490) and by the DFG (SFB1170). \end{acknowledgments}


\begin{thebibliography}{}
\bibitem{koch:natmater06}{S. W. Koch, M. Kira, G. Khitrova, and H. M. Gibbs, Nat. Mater. {\bf 5}, 523 (2006).}
\bibitem{mueller:2dma18}{T. Mueller and E. Malic, npj 2D Mater. Appl. {\bf 2}, 29 (2018).}
\bibitem{butov:jpcm04}{L. V. Butov, J. Phys.: Condens. Matter {\bf 16}, R1577 (2004).}
\bibitem{wang:nature19}{Z. Wang, D. A. Rhodes, K. Watanabe, T. Taniguchi, J. C. Hone, J. Shan, K. F. Mak, Nature {\bf 574}, 76 (2019).}
\bibitem{utsunomiya:natphys08}{S. Utsunomiya, L. Tian, G. Roumpos, C. W. Lai, N. Kumada, T. Fujisawa, M. Kuwata-Gonokami, A. L\"offler, S. H\"ofling, A. Forchel, and Y. Yamamoto, Nat. Phys. {\bf 4}, 700 (2008).}
\bibitem{byrnes:prl10}{T. Byrnes, T. Horikiri, N. Ishida, and Y. Yamamoto, Phys. Rev. Lett. {\bf 105}, 186402 (2010).}
\bibitem{lerario:natphys17}{G. Lerario, A. Fieramosca, F. Barachati, D. Ballarini, K. S. Daskalakis, L. Dominici, M. De Giorgi, S. A. Maier, G. Gigli, S. K\'ena-Cohen, D. Sanvitto, Nat. Phys. {\bf 13}, 837 (2017).}
\bibitem{fulde:pr64}{P. Fulde and R. A. Ferrell, Phys. Rev. {\bf 135}, A550 (1964).}
\bibitem{larkin:zetp64}{A. I. Larkin, Yu. N. Ovchinnikov, Zh. Eksp. Teor. Fiz. {\bf 47}, 1136 (1964).}
\bibitem{oldeolthof:prl21}{L. A. B. Olde Olthof, L. G. Johnsen, J. W. A. Robinson, and J. Linder, Phys. Rev. Lett. {\bf 127}, 267001 (2021).}
\bibitem{brydon:prl16}{P. M. R. Brydon, L. Wang, M. Weinert, and D. F. Agterberg, Phys. Rev. Lett. {\bf 116}, 177001 (2016).}
\bibitem{bahari:prr22}{M. Bahari, S.-B. Zhang, and B. Trauzettel, Phys. Rev. Research {\bf 4}, L012017 (2022).}
\bibitem{tang:prl21}{G. Tang, C. Bruder, and W. Belzig, Phys. Rev. Lett. {\bf 126}, 237001 (2021).}
\bibitem{hoerhold:2dmater23}{S. H\"orhold, J. Graf, M. Marganska, and M. Grifoni, 2D Mater. {\bf 10}, 025008 (2023).}
\bibitem{chakraborty:prb22}{D. Chakraborty and A. M. Black-Schaffer, Phys. Rev. B {\bf 106}, 024511 (2022).}
\bibitem{abrahams:prb95}{E. Abrahams, A. Balatsky, D. J. Scalapino, and J. R. Schrieffer, Phys. Rev. B {\bf 52}, 1271 (1995).}
\bibitem{lee:prb17}{S.-P. Lee, R. M. Lutchyn, and J. Maciejko, Phys. Rev. B {\bf 95}, 184506 (2017).}
\bibitem{parhizgar:prb21}{F. Parhizgar and A. M. Black-Schaffer, Phys. Rev. B {\bf 104}, 054507 (2021).}
\bibitem{moghaddam:prb14}{A. G. Moghaddam, T. Kernreiter, M. Governale, and U. Z\"ulicke, Phys. Rev. B {\bf 89}, 184507 (2014).}
\bibitem{kamide:prl10}{K. Kamide and T. Ogawa, Phys. Rev. Lett. {\bf 105}, 056401 (2010). }
\bibitem{lee:prl18}{K. H. Lee, C. Lee, H. Min, and S. B. Chung, Phys. Rev. Lett. {\bf 120}, 157601 (2018).}



	\bibitem{danovich:prb18}{M. Danovich, D. A. Ruiz-Tijerina, R. J. Hunt, M. Szyniszewski, N. D. Drummond, and V. I. Fal'ko, Phys. Rev. B {\bf 97}, 195452 (2018).}
	\bibitem{kogar:science17}{A. Kogar, M. S. Rak, S. Vig, A. A. Husain, F. Flicker, Y. I. Joe, L. Venema, G. J. MacDougall, T. C. Chiang, E. Fradkin, J. van Wezel, P. Abbamonte, Science {\bf 358}, 1314 (2017).}
	\bibitem{lozovik:prb02}{Yu. E. Lozovik, I. V. Ovchinnikov, S. Yu. Volkov, L. V. Butov, and D. S. Chemla, Phys. Rev. B {\bf 65}, 235304 (2002).}
	\bibitem{hanai:prb18}{R. Hanai, P. B. Littlewood, and Y. Ohashi, Phys. Rev. B {\bf 97}, 245302 (2018).}
	\bibitem{leon:prb01}{S. B. de-Leon and B. Laikhtman, Phys. Rev. B {\bf 63}, 125306 (2001).}
	\bibitem{kyriienko:prb12}{O. Kyriienko, E. B. Magnusson, and I. A. Shelykh, Phys. Rev. B {\bf 86}, 115324 (2012).}
	
	
	\bibitem{altland:book10}{A. Altland and B. D. Simons, {\it Condensed matter field theory}, (Cambridge University Press, Cambridge, 2010).}
	\bibitem{schwarz:natcommun20}{L. Schwarz, B. Fauseweh, N. Tsuji, N. Cheng, N. Bittner, H. Krull, M. Berciu, G. S. Uhrig, A. P. Schnyder, S. Kaiser, and D. Manske, Nat. Commun. {\bf 11}, 287 (2020).}
	\bibitem{chu:natcommun20}{H. Chu, M.-J. Kim, K. Katsumi, S. Kovalev, R. D. Dawson, L. Schwarz, N. Yoshikawa, G. Kim, D. Putzky, Z. Z. Li, H. Raffy, S. Germanskiy, J.-C. Deinert, N. Awari, I. Ilyakov, B. Green, M. Chen, M. Bawatna, G. Cristiani, G. Logvenov, Y. Gallais, A. V. Boris, B. Keimer, A. P. Schnyder, D. Manske, M. Gensch, Z. Wang, R. Shimano, and S. Kaiser, Nat. Commun. {\bf 11}, 1793 (2020).}
	\bibitem{matsunaga:prl13}{R. Matsunaga, Y. I. Hamada, K. Makise, Y. Uzawa, H. Terai, Z. Wang, and R. Shimano
Phys. Rev. Lett. {\bf 111}, 057002 (2013).}
\bibitem{mak:natnanotechnol12}{K. F. Mak, K. He, J. Shan, T. F. Heinz, Nat. Nanotechnol. {\bf 7}, 494 (2012).}
\bibitem{fogler:natcommun14}{M. M. Fogler, L. V. Butov, and K. S. Novoselov, Nat. Commun. {\bf 5}, 4555 (2014).}
\bibitem{shi:natnanotechnol22}{Q. Shi, E.-M. Shih, D. Rhodes, B. Kim, K. Barmak, K. Watanabe, T. Taniguchi, Z. Papi\'c, D. A. Abanin, J. Hone, and C. R. Dean, Nat. Nanotechnol. {\bf 17}, 577 (2022).}
\bibitem{calman:natcommun18}{E. V. Calman, M. M. Fogler, L. V. Butov, S. Hu, A. Mishchenko, A. K. Geim, Nat. Commun. {\bf 9}, 1895 (2018).}
\bibitem{cai:nanolett24}{C.-S. Cai, W.-Y. Lai, P.-H. Liu, T.-C. Chou, R.-Y. Liu, C.-M. Lin, S. Gwo, and W.-T. Hsu, Nano Lett. {\bf 24}, 2773 (2024).}
\bibitem{trainer:acs20}{D. J. Trainer, B. K. Wang, F. Bobba, N. Samuelson, X. Xi, J. Zasadzinski, J. Nieminen, A. Bansil, and M. Iavarone, ACS Nano {\bf 14}, 2718 (2020).}
\bibitem{luepke:natphys20}{F. L\"upke, D. Waters, S. C. de la Barrera, M. Widom, D. G. Mandrus, J. Yan, R. M. Feenstra, and B. M. Hunt, Nat. Phys. {\bf 16}, 526 (2020).}
\bibitem{hasling:apl15}{M. W. Hasling, Y. Y. Kuznetsova, P. Andreakou, J. R. Leonard, E. V. Calman, C. J. Dorow, M. Hanson, A. C. Gossard, Appl. Phys. Lett. {\bf 117}, 023108 (2015).}
\bibitem{bopp:prb23}{F. Bopp, J. Schall, N. Bart, F. V\"ogl, C. Cullip, F. Sbresny, K. Boos, C. Thalacker, M. Lienhart, S. Rodt, D. Reuter, A. Ludwig, A. D. Wieck, S. Reitzenstein, K. M\"uller, and J. J. Finley, Phys. Rev. B {\bf 107}, 165426 (2023).}
\bibitem{cheng:prb00}{H. H. Cheng, S. T. Yen, and R. J. Nicholas, Phys. Rev. B {\bf 62}, 4638 (2000).}
\bibitem{mayer:aem20}{W. Mayer, W. F. Schiela, J. Yuan, M. Hatefipour, W. L. Sarney, S. P. Svensson, A. C. Leff, T. Campos, K. S. Wickramasinghe, M. C. Dartiailh, I. Zutic, and J. Shabani, ACS Appl. Electron. Mater. {\bf 2}, 2351 (2020).}
\bibitem{skrotzki:apl10}{R. Skrotzki, J. Fiedler, T. Herrmannsd\"orfer, V. Heera, M. Voelskow, A. M\"ucklich, B. Schmidt, W. Skorupa, G. Gobsch, M. Helm, and J. Wosnitza, Appl. Phys. Lett. {\bf 97}, 192505 (2010).}
\end{thebibliography}
\end{document}